\documentstyle[12pt,twoside,fleqn,espcrc1,epsbox]{article}

\newcommand{\AmS}{{\protect\the\textfont2
  A\kern-.1667em\lower.5ex\hbox{M}\kern-.125emS}}
\newcommand{\beq}{\begin{eqnarray}}
\newcommand{\eeq}{\end{eqnarray}}

\title{$A_{LT}$ in the Nucleon-Nucleon Polarized Drell-Yan Process}

\author{Y. Kanazawa\address{Graduate School 
of Science and Technology, Niigata University, Nigata 950-2181, Japan}, 
Yuji Koike\address{Department of Physics, Niigata University,
Niigata 950-2181, Japan} and N. Nishiyama$^{\rm a}$
}

\begin{document}
\maketitle

\begin{abstract}
We present a leading order (LO) 
estimate for the longitidinal-transverse 
spin asymmetry ($A_{LT}$) in the nucleon-nucleon polarized
Drell-Yan process at RHIC and HERA-$\vec{N}$ energies
in comparison with $A_{LL}$ and $A_{TT}$.
$A_{LT}$ receives contribution from $g_1$, the transversity
distribution $h_1$, and the twist-3 distributions $g_T$
and $h_L$.  For the twist-3 contribution we use
the bag model prediction evolved to a high energy scale by
the large-$N_c$ evolution equation.  We found that
$A_{LT}$ (normalized by the asymmetry in the parton level)
is much smaller than the corresponding $A_{TT}$.
Twist-3 contribution given by the bag model 
turned out to be negligible.
\end{abstract}

\vspace{0.5cm}

The nucleon-nucleon scattering provides us with a new opportunity
to probe nucleon's internal structure.
In particular, polarized Drell-Yan
lepton pair production opens a window toward
new types of spin dependent parton distributions -- chiral-odd
distributions $h_1(x,\mu^2)$
and $h_L(x,\mu^2)$ which can not be measured by the deep
inelastic lepton-nucleon scatterings\,\cite{RS}.
There are 
three kinds of double
spin asymmetries in the nucleon-nucleon polarized Drell-Yan process:
They are $A_{LL}$ (collision between the  
longitudinally polarized nucleons), $A_{TT}$ (
collision between the 
transversely polarized nucleons), and $A_{LT}$ (longitudinal
versus transverse).
The experimental data on these asymmetries will presumably
be reported by RHIC at BNL and HERA-$\vec{N}$ at DESY.
By now
several reports are already available for the
estimate of $A_{LL}$ and
$A_{TT}$
in the next-to-leading order (NLO)
level\,\cite{DY}. 
In this talk we present
a first
estimate on $A_{LT}$ in comparison with $A_{LL}$ and $A_{TT}$
at RHIC and HERA energies in the LO QCD\,\cite{KKN}. 
$A_{LT}$ is particularly interesting, since
it receives the twist-3 contribution as a leading contribution
(although it is proportional to $1/Q$), 
giving a possibility
of seeing quark-gluon correlation in hard processes.

In LO QCD, the double spin asymmetries
are given by
\beq
A_{LL} &=& \frac{\sigma (+,+)-\sigma (+,-) }
                {\sigma (+,+)+\sigma (+,-) }
       =  \frac{\Sigma_{a} e_{a}^2
                   g_{1}^a(x_1,Q^2)g_{1}^{\bar{a}}(x_2,Q^2)}
               {\Sigma_{a} e_{a}^2 f_{1}^a(x_1,Q^2)f_{1}^{\bar{a}}(x_2,Q^2)},
\label{ALL}\\[5pt]
A_{TT} &=& \frac{\sigma (\uparrow,\uparrow)-\sigma (\uparrow,\downarrow) }
                {\sigma (\uparrow,\uparrow)+\sigma (\uparrow,\downarrow) }
       = a_{TT} \frac{\Sigma_{a} e_{a}^2
                   h_{1}^a(x_1,Q^2)h_{1}^{\bar{a}}(x_2,Q^2)}
               {\Sigma_{a} e_{a}^2 f_{1}^a(x_1,Q^2)f_{1}^{\bar{a}}(x_2,Q^2)},
\label{ATT}\\[5pt]
A_{LT} &=& \frac{\sigma (+,\uparrow)-\sigma (+,\downarrow) }
                {\sigma (+,\uparrow)+\sigma (+,\downarrow) }
       = a_{LT} \frac{\Sigma_{a} e_{a}^2
                   \left [g_{1}^a(x_1,Q^2)x_2g_{T}^{\bar{a}}(x_2,Q^2)
                        + x_1h_{L}^a(x_1,Q^2)h_{1}^{\bar{a}}(x_2,Q^2) \right]}
               {\Sigma_{a} e_{a}^2 f_{1}^a(x_1,Q^2)f_{1}^{\bar{a}}(x_2,Q^2)},
\nonumber\\
\label{ALT}
\end{eqnarray}
where $\sigma(S_1,S_2)$ represents the Drell-Yan cross section
with the two nucleon's spin $S_1$ and $S_2$, 
$e_a$ represent the electric charge of the quark-flavor
$a$ and the summation is over all quark and anti-quark flavors: 
$a=u,d,s,\bar{u},\bar{d},\bar{s}$, ignoring heavy quark 
contents ($c,b,\cdots$) 
in the nucleon.
The variables $x_1$ and $x_2$ refer to
the momentum fractions of the partons coming from
the two nucleons ``1'' and ``2'', respectively.
In (\ref{ATT}) and (\ref{ALT}), $a_{TT}$ and $a_{LT}$ represent
the asymmetries in the parton level defined as
$a_{TT}  =  {\rm sin}^2\theta\, 
{\rm cos}2\phi/(1+{\rm cos}^2\theta)$ and 
$a_{LT}  =  (M/Q) (2\,{\rm sin}2\theta\, 
{\rm cos}\phi)/(1+{\rm cos}^2\theta)$,
where $\theta$ and $\phi$ are, respectively, the
polar and azimuthal angles of the virtual photon in the center of mass system 
with respect to the beam direction and 
the transverse spin.
We note that $A_{LL}$ and $A_{TT}$ receive contribution only from
the twist-2 distributions, while $A_{LT}$ is proportional to the twist-3 
distributions
and hence $a_{LT}$ is suppressed by a factor $1/Q$.

The twist-3 distributions $g_T$ and $h_L$ can be decomposed into
the twist-2 contribution  
and the ``purely twist-3'' contribution:
\beq
g_{T}(x,\mu^2) = \int_{x}^{1} dy\frac{g_{1}(y,\mu^2)}{y}  + 
\widetilde{g}_{T}(x,\mu^2);\quad
h_{L}(x,\mu^2) = 2x\int_{x}^{1}dy \frac{h_{1}(y,\mu^2)}{y^2}  
+ \widetilde{h}_{L}(x,\mu^2).
\label{gTWW}
\eeq
The purely twist-3 pieces
$\widetilde{g}_T$ and
$\widetilde{h}_L$ can be written as 
quark-gluon-quark correlators on the lightcone.
In the following we call the 
first terms in (\ref{gTWW}) 
$g_T^{WW}(x,\mu^2)$ and $h_L^{WW}(x,\mu^2)$ (Wandzura-Wilczek parts).

For the present estimate of $A_{LT}$,
we use 
the LO parametrization for $f_1$ by 
Glu\"ck-Reya-Vogt\,\cite{GRV} and
the
LO parametrization (standard scenario) for $g_1$
by Glu\"ck-Reya-Stratmann-Vogelsang (GRSV)\,\cite{GRSV}.
For $h_1$, $g_T$ and $h_L$ no experimental
data is available up to now and we have to rely on some theoretical
postulates. 
Here we assume $h_1(x,\mu^2)=g_1(x,\mu^2)$ at a low energy scale
($\mu^2=0.23$ GeV$^2$) 
as has been suggested by a low energy nucleon model\,\cite{JJ}.
These assumptions also fix $g_T^{WW}$ and $h_L^{WW}$.
For the purely twist-3 parts $\widetilde{g}_T$ and
$\widetilde{h}_L$ we employ the bag model results at a low energy 
scale, assuming the bag scale is $\mu_{bag}^2=0.081$ and $0.25$ GeV$^2$.  
In particular, we set the strangeness contributions
to the purely twist-3 contributions equal to zero.
By these boundary conditions for $h_1$, $g_T$ and $h_L$
at a low energy side
and applying the relevant $\mu^2$ evolution to them,
we can estimate $A_{LT}$. 

The $\mu^2$ evolution of 
the twist-3 distributions is 
quite comlicated\,\cite{KT}.
However, 
it has been proved that at large $N_c$ their $\mu^2$-dependence 
can be described by a simple DGLAP evolution equation
similarly to the twist-2 distributions 
and the correction due to the finite value of $N_c$ is 
of $O(1/N_c^2)\sim 10$ \% level\,\cite{ABH}.
Here we apply this large-$N_c$
evolution to the bag model results\,\cite{KK}.

The double spin asymmetries are the functions of
the square of the center-of-mass energy
$s=(P_1 + P_2)^2$ ($P_1$ and $P_2$ are the four momenta of the
two nucleons), 
the squared invariant mass of
the lepton pair 
$Q^2=(x_1 P_1 + x_2 P_2)^2 = x_1 x_2 s$ ($M^2 << Q^2$) and the Feynman's 
$x_F=2q_3/ \sqrt{s}=x_1-x_2$.  
Using these variables, momentum fractions of
each quark and anti-quark in (\ref{ALL})-(\ref{ALT}) can be written
as
$x_1 = \left( x_F +\sqrt{x_F^2 + (4Q^2/s)}\right)/2$ and 
$x_2 = \left( -x_F +\sqrt{x_F^2 + (4Q^2/s)}\right)/2$.

Figure 1 shows the three asymmetries normalized by
the asymmetries in the parton level, $\widetilde{A}_{LL}=-A_{LL}$,
$\widetilde{A}_{TT}=-A_{TT}/a_{TT}$, $\widetilde{A}_{LT}=-A_{LT}/a_{LT}$.  
They are plotted as a function of $x_F$
for fixed values of 
$Q=\sqrt{Q^2}$ ($=8,\ 10$ GeV) and $\sqrt{s}$ ($=50,\ 200$ GeV),
which are within or 
close to the planned RHIC and HERA-$\vec{N}$
kinematics. ($50\ {\rm GeV}< \sqrt{s}< 500\ {\rm GeV}$ 
for RHIC,
and $\sqrt{s}=39.2$ GeV for HERA-$\vec{N}$.)  
$\widetilde{A}_{LL}$ and 
$\widetilde{A}_{TT}$ are symmetric with respect to $x_F=0$, while
$\widetilde{A}_{LT}$ is not symmetric 
as is obvious from the kinematics.  In general all these asymmetries are
larger for larger $Q^2/s$.
$\widetilde{A}_{LT}$ with only the twist-2 contributions in $g_T$ and $h_L$
are shown by solid lines.
They are
typically 5 to 10 times smaller than $\widetilde{A}_{LL}$ and
$\widetilde{A}_{TT}$.  
$\widetilde{A}_{LT}$ with complete $g_T$ and $h_L$ is shown by
the short dash-dot ($\mu_{bag}^2=0.25$ GeV$^2$) and the dotted
($\mu_{bag}^2=0.081$ GeV$^2$) lines.  
Since large $|x_F|$ corresponds to small $x_1$ or $x_2$,
and the bag model prediction for the distribution function
becomes unreliable in the small-$x$ region, we only plotted these lines
for the region $x_1, x_2 > 0.07$.   As can be seen from
Fig. 1, the purely twist-3 contribution brings only tiny correction
to $\widetilde{A}_{LT}$.   Larger value of the bag scale
$\mu^2_{bag}$ would not make it appreciably larger.  
The smallness of $\widetilde{A}_{LT}$
can be ascribed to the factors $x_1$ or $x_2$ in (\ref{ALT}).
In the kinematic range considered either $x_1$ or $x_2$ (or both)
take very small values.  
If it were not for those factors, $\widetilde{A}_{LT}$
would be comparable to $\widetilde{A}_{LL}$ and $\widetilde{A}_{TT}$.  
We remind in passing 
that what is measured experimentaly is $A_{LT}$ itself
which receives the suppression factor $M/Q$ from $a_{LT}$.

To summarize, we presented a first estimate of the
longitudinal-transverse spin asymmetry $A_{LT}$ 
for the polarized Drell-Yan process at RHIC and HERA-$\vec{N}$ energies
in comparison with $A_{LL}$ and $A_{TT}$.  
$A_{LT}$ normalized by the asymmetry in the parton level 
turned out to be approximately five to ten times smaller than
the corresponding $A_{TT}$, although the prediction on its absolute 
magnitude suffers from the uncertainty of the distributions, 
in particular, 
of $h_1$ as was the case for $A_{TT}$.  
The purely twist-3 contribution to $g_T$ and $h_L$ was modeled
by the bag model, and it turned out its effect on 
$A_{LT}$ is negligible compared with
the Wandzura-Wilczek contribution to $g_T$ and $h_L$.

\begin{figure}[h]
\epsfile{file=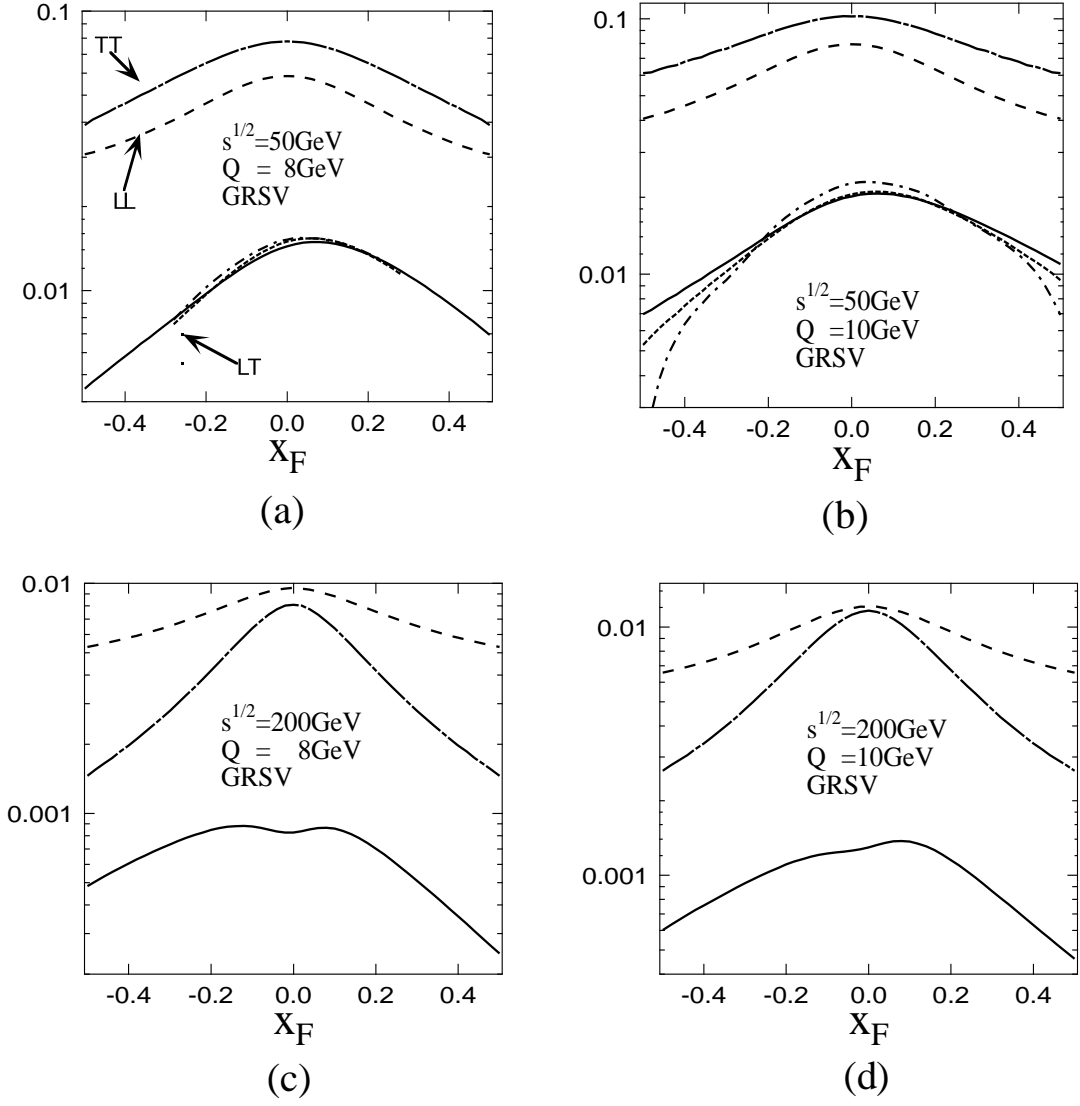,scale=0.85}
\caption[]{Double spin asymmetries, $\widetilde{A}_{LL}$,
$\widetilde{A}_{TT}$, $\widetilde{A}_{LT}$, 
for the polarized Drell-Yan
using the GRSV parton distribution and the bag model at
$Q=8,\ 10$ GeV and $\sqrt{s}=50,\ 200$ GeV.
The solid line denotes $\widetilde{A}_{LT}$ with only the
Wandzura-Wilczek contributions in $g_T$ and $h_L$. 
The short dash-dot line denotes $\widetilde{A}_{LT}$ with 
the bag scale
$\mu_{bag}^2=0.25$ GeV$^2$, and the dotted line
denotes
$\widetilde{A}_{LT}$ with the bag scale
$\mu_{bag}^2=0.081$ GeV$^2$.
The long dashed line corresponds to $\widetilde{A}_{LL}$, 
and the long dash-dot line corresponds to $\widetilde{A}_{TT}$.}
\end{figure}

\end{document}